\documentclass[prx,letterpaper,nobalancelastpage,twocolumn,superscriptaddress,nofootinbib]{revtex4-2}

\usepackage{graphicx}
\usepackage{amsmath}
\usepackage{bbold}
\usepackage{amssymb}
\usepackage[english]{babel}
\usepackage{color}
\usepackage[version=4]{mhchem}
\usepackage[hidelinks]{hyperref}
\usepackage{dsfont}
\usepackage{placeins}
\usepackage{mathtools}
\usepackage[normalem]{ulem}
\usepackage{lipsum}

\setcitestyle{super}

\usepackage{soul}
\newcommand{\ket}[1]{| #1 \rangle}
\newcommand{\bra}[1]{\langle #1 |}

\newcommand{\Caltech}{California Institute of Technology, Pasadena, CA 91125, USA}

\newcommand{\Stanford}{Department of Electrical Engineering, Stanford University, Stanford, CA, USA}

\usepackage{caption}

\DeclareCaptionLabelSeparator{bar}{ \textbf{\textbar}~}
\usepackage{enumitem}
\setlist{nolistsep}
\begin{document}

\title{Erasure conversion in a high-fidelity Rydberg quantum simulator}

\author{Pascal Scholl}\thanks{These authors contributed equally to this work}
\author{Adam L. Shaw}\thanks{These authors contributed equally to this work}
\author{Richard Bing-Shiun Tsai}
\author{Ran Finkelstein}
\affiliation{\Caltech}
\author{\\Joonhee Choi}
\affiliation{\Caltech}
\affiliation{\Stanford}
\author{Manuel Endres}\email{mendres@caltech.edu}
\affiliation{\Caltech}

\maketitle

\textbf{Minimizing and understanding errors is critical for quantum science, both in noisy intermediate scale quantum (NISQ) devices~\cite{Preskill2018} and for the quest towards fault-tolerant quantum computation~\cite{Shor1995,Knill1996}. Rydberg arrays have emerged as a prominent platform in this context~\cite{Saffman2016} with impressive system sizes~\cite{Scholl2021,Ebadi2021} and proposals suggesting how error-correction thresholds could be significantly improved by detecting leakage errors with single-atom resolution\cite{Wu2022,Sahay2023}, a form of erasure error conversion~\cite{Grassl1997,Kang2023,Teoh2022,Kubica2022}. However, two-qubit entanglement fidelities in Rydberg atom arrays~\cite{Madjarov2020, Levine2019} have lagged behind competitors~\cite{Clark2021,Negirneac2021} and this type of erasure conversion is yet to be realized for matter-based qubits in general. Here we demonstrate both erasure conversion and high-fidelity Bell state generation using a Rydberg quantum simulator~\cite{Bernien2017,Ebadi2021,Scholl2021,Choi2023}. When excising data with erasure errors observed via fast imaging of alkaline-earth atoms~\cite{Cooper2018,Norcia2018,Saskin2019,Jenkins2022}, we achieve a Bell state fidelity of ${\geq} 0.9971^{+10}_{-13}$, which improves to ${\geq}0.9985^{+7}_{-12}$ when correcting for remaining state preparation errors. We further apply erasure conversion in a quantum simulation experiment for quasi-adiabatic preparation of long-range order across a quantum phase transition, and unveil the otherwise hidden impact of these errors on the simulation outcome. Our work demonstrates the capability for Rydberg-based entanglement to reach fidelities in the ${\sim} 0.999$ regime, with higher fidelities a question of technical improvements, and shows how erasure conversion can be utilized in NISQ devices. The shown techniques could be translated directly to quantum error-correction codes with the addition of long-lived qubits~\cite{Burgers2022,Jenkins2022,Ma2022,Wu2022}.
}

\begin{figure*}[t!]
	\centering
	\includegraphics[width=\textwidth]{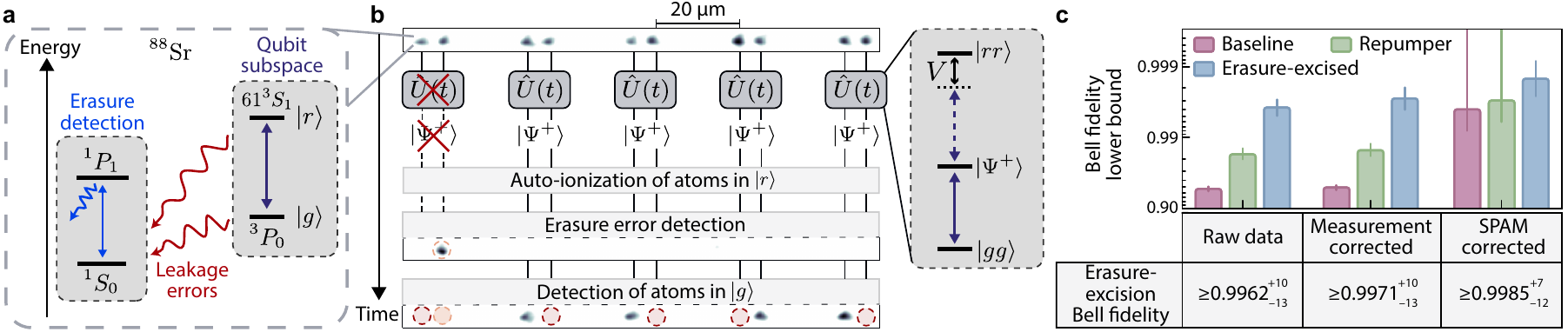}
	\caption{\textbf{Erasure conversion for high-fidelity entanglement.} \textbf{a,} Level structure used in this work. We distinguish two subspaces: a qubit subspace in which the atoms interact via their Rydberg states, and a measurement subspace used to detect leakage errors from the qubit subspace with single-site resolution, realizing erasure conversion. \textbf{b,} Sketch of the erasure conversion scheme, as applied to Bell pair generation. After arranging atoms into pairs (top) we prepare them in $\ket{g}$, and entangle them via the Rydberg blockade mechanism (right), denoted by a unitary operation $\hat{U}(t)$. Immediately afterwards, we auto-ionize atoms in $\ket{r}$, effectively projecting the populations of the Bell states, and follow with a fast erasure conversion image to detect leakage out of the qubit subspace during the preparation or evolution periods. This is followed by the final detection of atoms in $\ket{g}$, yielding two separate, independent images. We can discard data from pairs where atoms are detected in the erasure-error image, termed erasure-excision in the following. Atom fluorescence images are single-shot, with post-processing applied to improve detection fidelity~\cite{Bergschneider2018} (Methods). \textbf{c,} Lower bounds for Bell state fidelities with (blue) and without (red) the erasure-excision, and using incoherent pumping to reduce preparation errors instead of erasure-excision (green, Methods). We present the results for the raw data, corrected for measurement errors, and corrected for state preparation and measurement errors (SPAM). All data is averaged over eight pairs of atoms which are excited in parallel. Error bars represent a $68\%$ confidence interval (Ext. Data Fig.~\ref{EFig2}, Methods).
 }
	\vspace{-0.5cm}
	\label{Fig1}
\end{figure*}

We begin by detailing our erasure conversion scheme and how it is employed in conjunction to Bell state generation, resulting in fidelities competitive with other state-of-the-art platforms~\cite{Clark2021,Negirneac2021,Bruzewicz2019,Kjaergaard2020}. Our experimental apparatus has been described in detail before~\cite{Madjarov2020}, and is based on trapping individual strontium atoms in arrays of optical tweezers~\cite{Cooper2018,Norcia2018} (Methods). Strontium features a rich energy structure, allowing us to utilize certain energy levels as a qubit subspace to perform entangling operations and separate levels for detection of leakage errors (Fig.~\ref{Fig1}a).

To controllably generate entanglement between atoms, we employ Rydberg interactions~\cite{Lukin2001, Gaetan2009, Isenhower2010}. When two atoms in close proximity are simultaneously excited to high-lying electronic energy levels, called Rydberg states, they experience a distance-dependent van der Waals interaction $V=C_6/r^6$, where $r$ is the interatomic spacing, and $C_6$ is an interaction coefficient. If the Rabi frequency, $\Omega$, which couples the ground, $\ket{g}$, and Rydberg, $\ket{r}$, states is much smaller than the interaction shift, $\Omega/V \ll 1$, the two atoms cannot be simultaneously excited to the Rydberg state (Fig.~\ref{Fig1}b, inset), a phenomena known as Rydberg blockade. In this regime, the laser drives a unitary operation, $\hat{U}(t)$, that naturally results in the two atoms forming a Bell state, $\ket{\Psi^+} = \frac{1}{\sqrt{2}}(\ket{gr}+\ket{rg})$, between the ground and Rydberg states (Fig.~\ref{Fig1}b). 

This Bell state generation has several major practical limitations. Of particular interest here are leakage errors to the absolute ground state, ${}^1S_0$, which are converted to erasure errors in our work as described below (and in Ext. Data Fig.~\ref{EFig1}). The first error of this type is imperfect preparation of atoms in $\ket{g}$ prior to applying $\hat{U}(t)$. The second arises from decay out of the Rydberg state along multiple channels. We distinguish decay into `bright' states, which we can image, and `dark' states, which are undetected (Ext. Data Fig.~\ref{EFig_decay}). The former primarily refers to low-lying energy states which are repumped to ${}^1S_0$ as part of the imaging process or decay to ${}^1S_0$ via intermediate states, while the latter mainly consists of nearby Rydberg states accessed via blackbody radiation.

Here we employ a scheme, theoretically proposed~\cite{Wu2022} but not yet demonstrated, which allows us to detect the location of such leakage errors (Fig.~\ref{Fig1}b), converting them into so-called erasure errors, i.e., errors with a known location~\cite{Grassl1997}. To this end, we demonstrate fast, 24 $\mu$s imaging of atoms in ${}^1S_0$ (Ext. Data. Fig.~\ref{EFig1}) with single-site resolution and $0.980^{+1}_{-1}$ fidelity. Such fast imaging had previously been performed for a few, freely-propagating, alkali atoms~\cite{Bergschneider2018}, but not for many trapped atoms in tweezer arrays or alkaline-earth atoms (Methods). 

Our general procedure is shown in Fig.~\ref{Fig1}b (further detailed in Ext. Data Fig.~\ref{EFig_sequence}). We first rearrange~\cite{Endres2016,Barredo2016} atoms into pairs, coherently transfer them to $\ket{g}$, and then perform the entangling $\hat{U}$ operation. Immediately after, we auto-ionize the atoms to project the populations of the resultant state. 

We then perform the fast erasure image; any atoms which are detected are concluded to be the result of some leakage error process. Importantly, the erasure image does not affect atoms remaining in $\ket{g}$, and is extremely short compared to its lifetime, resulting in a survival probability in $\ket{g}$ of $0.9999954^{+12}_{-12}$ (Ext. Data Fig.~\ref{EFig1}, Methods). Hence, the erasure image does not perturb the subsequent final readout. Thus, we obtain two separate images characterizing a single experimental repetition, the final image showing the ostensible result of $\hat{U}$, and the erasure image revealing leakage errors with single-site resolution.

We note that this work is not a form of mid-circuit detection as no superposition states of $\ket{g}$ and $\ket{r}$ exist at the time of the erasure image. Instead, our approach is a noise mitigation strategy via erasure-excision, where experimental realizations are discarded if erasures are detected. In contrast to other leakage mitigation schemes previously demonstrated in matter-based qubit platforms~\cite{Hayes2020,Stricker2020,McEwen2021}, we directly spatially resolve leakage errors in a way which is decoupled from the performed experiment, is not post-selected on the final qubit readout, and does not require any extra qubits to execute. 

However, the coherence between $\ket{g}$ and $\ket{r}$ can in principle be preserved during erasure detection for future applications; in particular, we see no significant difference in Bell state lifetime with and without the imaging light for erasure detection on (Ext. Data Fig.~\ref{EFig_coherence}, Methods). We also expect long-lived nuclear qubits encoded in $\ket{g}$ to be unperturbed by our implementation of erasure conversion~\cite{Burgers2022,Jenkins2022,Ma2022,Wu2022}.

\vspace{0.25cm}
\noindent\textbf{Bell state generation results}\newline
With a procedure for performing erasure conversion in hand, we now describe its impact on Bell state generation. Experimentally, we only obtain a lower-bound for the Bell state generation fidelity~\cite{Madjarov2020} (Methods, Ext. Data Fig.~\ref{EFig2}); the difference of this lower bound to the true fidelity is discussed further below.

We first coherently transfer atoms to $\ket{g}$ as described before, and then consider three scenarios (Fig.~\ref{Fig1}c, Ext. Data Table~\ref{EFig:table_Bell}). In the first, as a baseline we perform the entangling unitary $\hat{U}$ without considering any erasure detection results (red bars). In the second, we excise data from any pairs of atoms with an observed erasure error (blue bars). Finally, we compare against another strategy for mitigating preparation errors through incoherent repumping~\cite{Madjarov2020}, but without erasure detection (green bars). Notably, the raw value for the Bell state lower-bound with erasure-excision is ${\geq} 0.9962^{+10}_{-13}$,  significantly higher than with the other methods. This difference mainly comes from erasure excision of preparation errors and, to a much lower degree, Rydberg decay. These contribute at the level of ${\sim} 5\times10^{-2}$ and $1.2^{+3}_{-3}\times10^{-4}$, respectively (Methods).

Correcting for final measurement errors, we find a lower bound of ${\geq} 0.9971^{+10}_{-13}$, which quantifies our ability to generate Bell pairs conditioned on finding no erasure events. To quantify the quality of the Rydberg entangling operation $\hat{U}(t)$ itself, we further correct for remaining preparation errors that are not detected in the erasure image (Methods), and find a state preparation and measurement (SPAM) corrected lower bound of ${\geq} 0.9985^{+7}_{-12}$.

To our knowledge, these bare, measurement-corrected, and SPAM-corrected values are respectively the highest two-qubit entanglement fidelities measured for neutral atoms to date, independent of the means of entanglement generation. While Bell state generation as demonstrated here is not a computational two-qubit quantum gate -- which requires additional operations - our results are indicative of the fidelities achievable in Rydberg based gate operations.

\begin{figure}[t!]
	\centering
	\includegraphics[width=\columnwidth]{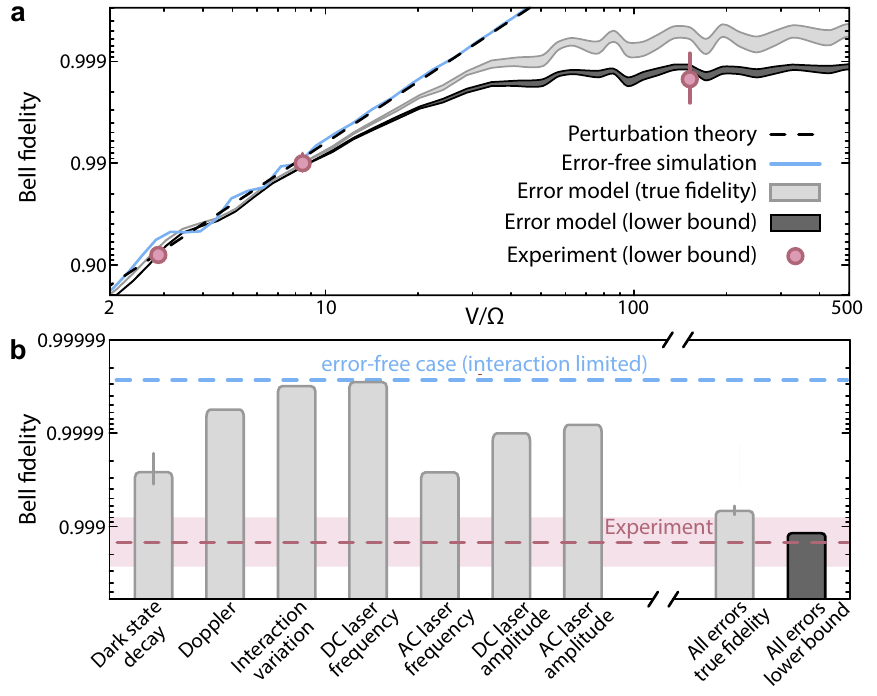}
	\caption{\textbf{
Predicting infidelities at the $10^{-3}$ level.} \textbf{a,} SPAM-corrected Bell state fidelity as a function of the ratio of interaction energy and Rabi frequency, $V/\Omega$. Error-free simulations (blue line) show fidelities continually increase with increasing $V/\Omega$, in agreement with results from perturbation theory (dashed line). For large enough interaction strength ($V/\Omega>50$), other error sources become dominant, and we employ a noisy open system dynamics simulation from which we obtain an estimate of the true fidelity (light grey fill) and for the lower bounding procedure used in experiment (dark grey fill). We find good agreement between simulation and experimental results with erasure-excision (red markers). \textbf{b,} Predicted Bell state fidelity for $V/\Omega = 140$ from simulations turning on a single noise term at a time. Dominant limitations come from laser frequency and intensity noise, as well as decay of the Rydberg state into dark states. We also show the results when taking into account all errors (Methods), for both the true fidelity and the lower bound estimation (right). The lower-bound significantly underestimates the true fidelity. Shaded areas in \textbf{a} and error bars in \textbf{b} represent the standard deviation of the mean over 5000 trajectories.
 }
	\vspace{-0.5cm}
	\label{Fig2}
\end{figure}

\vspace{0.25cm}
\noindent\textbf{Error modelling}\newline
Importantly, we understand remaining errors in the entangling operation as well the nature of detected erasure errors from a detailed \textit{ab-initio} error model simulation for SPAM-corrected fidelities (Methods, Fig.~\ref{Fig2}). We identify limited interaction strength as a dominant effect that restricted SPAM-corrected entanglement fidelities in our previous work~\cite{Madjarov2020} (Fig.~\ref{Fig2}a); in particular, one major difference here is that we operate at smaller distance and hence larger $V/\Omega$. In line with experimental data (red markers), fidelities at large distances are limited to $F_\textrm{Bell} \leq 1-\frac{5}{8}(\Omega/V)^2$ obtained from perturbation theory (black dashed line, Methods).

For strong enough interaction, $V/\Omega>50$, corresponding to distances $r<3\, \mu\text{m}$, other error sources become limiting. In this short-distance regime, the experimental SPAM-corrected fidelity lower-bound is in good agreement with the error model prediction of ${\geq}0.99881^{+3}_{-3}$ (dark grey fill). 

Our error model results show that the lower bound procedure significantly underestimates the true fidelity (light grey fill), found to be $0.99931^{+6}_{-6}$. This effect arises because the lower bound essentially evaluates the fidelity of $\hat{U}$ by a measurement after performing $\hat{U}$ twice (Methods), meaning particular errors can be exaggerated. Given the good match of the error model and experimental fidelity lower bounds, we expect this effect to be present in experiment as well, and to underestimate the true SPAM-corrected fidelity by about $5\times10^{-4}$.

The remaining infidelity is a combination of multiple errors. In Fig.~\ref{Fig2}b, we report an error budget for the most relevant noise source contributions to the Bell state infidelity (Methods) at the experimentally chosen $V/\Omega=140$. Frequency and intensity laser noise are dominant limitations, but could be alleviated by improving the stability of laser power, and reducing its linewidth, for instance via cavity filtering~\cite{Levine2018}. Eliminating laser noise completely would lead to fidelities of ${\sim}0.9997$ in our model. The other major limit is Rydberg state decay into dark states, which cannot be converted into an erasure detection with our scheme. This decay is mostly blackbody induced~\cite{Wu2022,Low2012}, and thus could be greatly reduced by working in a cryogenic environment~\cite{Schymik2021}, leaving mostly spontaneous decay that is bright to our erasure detection. Accounting for these improvements, it is realistic that Rydberg-based Bell state generation in optical tweezers arrays could reach ${>}0.9999$ fidelity in the coming years.

\begin{figure*}[t!]
	\centering
	\includegraphics[width=\textwidth]{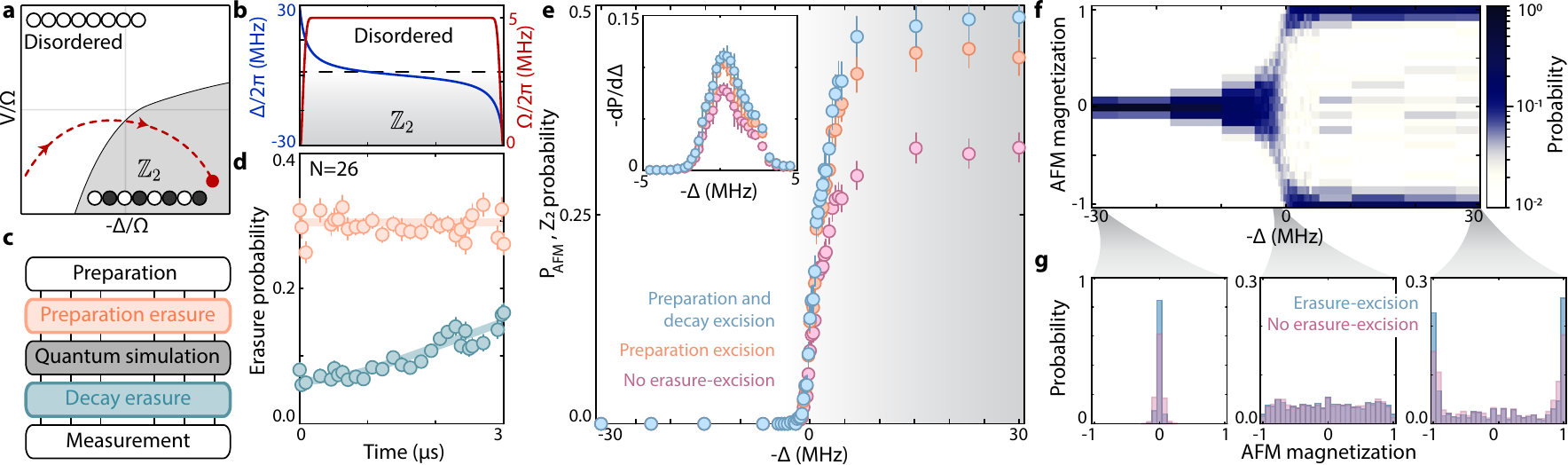}
	\caption{\textbf{Erasure conversion in quantum simulation.} \textbf{a, b,} We perform quasi-adiabatic sweeps with $N=26$ spins in the effective ground state manifold of an attractive Rydberg Hamiltonian (Methods), starting from the initially disordered phase, and ending in the $\mathbb{Z}_2$ ordered phase. \textbf{c,} We perform two erasure images, one preceding the quantum simulation (to detect preparation errors), and one following (to partially detect Rydberg decay errors). \textbf{d,} The probability for detecting a preparation error (orange markers) remains constant while the probability for detecting a decay error (green markers) grows in agreement with the Rydberg lifetime and detection infidelities (solid lines, Methods). \textbf{e,} The total probability, $P_\textrm{AFM}$, for forming either of the AFM $\mathbb{Z}_2$ states is improved by performing erasure-excision on all errors (blue markers), as compared to only on preparation errors (orange markers) or performing no excision (pink markers). The sensitivity of $P_\textrm{AFM}$ with respect to a change in $\Delta$ also increases with erasure-excision (inset). \textbf{f,} The probability distribution for measuring a given AFM magnetization is initially peaked at 0 in the disordered phase, before bifurcating when entering the $\mathbb{Z}_2$ phase, consistent with spontaneous symmetry breaking. \textbf{g,} Deep in either phase, erasure-excision leads to a sharpening of the probability distribution (left, right). Around the phase transition, we observe a close-to-flat distribution (middle).
 }
	\vspace{-0.5cm}
	\label{Fig3}
\end{figure*}

\vspace{0.25cm}
\noindent\textbf{Quantum simulation with erasure conversion}\newline
Having demonstrated the benefits of erasure-excision for the case of improving two-qubit entanglement fidelities, we now show it can be similarly applied to the case of many-body quantum simulation, demonstrating the utility of erasure detection for NISQ applications. As part of this investigation, we also distinguish erasure errors from preparation and Rydberg spontaneous decay, the latter of which becomes more visible in a many-body setting and for longer evolution times.

As a prototypical example, we explore a quasi-adiabatic sweep into a $\mathbb{Z}_2$-ordered phase (Fig.~\ref{Fig3}a) through the use of a varying global detuning~\cite{Fendley2004} (Fig.~\ref{Fig3}b). In this ordered phase, ground and Rydberg states form an antiferromagnetic (AFM) pattern, with long-range order appearing at a quantum phase transition. Unlike previous examples~\cite{Bernien2017,Omran2019}, we operate in the effectively \textit{attractive} interacting regime of the Rydberg blockaded space~\cite{Fendley2004}, which features a true two-fold degenerate ground state for systems with an even number of atoms, even for open boundary conditions (Methods), and without explicitly modifying the boundary~\cite{Omran2019}. The ground state in the deeply ordered limit consists of two oppositely ordered AFM states, $\ket{grgr...gr}$ and $\ket{rgrg...rg}$. 

Staying adiabatic during ground state preparation requires evolution over microseconds, orders of magnitude longer than the two-qubit entanglement operation shown before, which magnifies the effect of Rydberg decay. In order to differentiate between leakage out of the qubit manifold due to either preparation errors or Rydberg decay, we perform two erasure images, one before the adiabatic sweep which captures preparation errors, and one after (Fig.~\ref{Fig3}c). The second image allows us to measure Rydberg decay into the detection subspace throughout the sweep. For a system size of $N=26$ atoms (Fig.~\ref{Fig3}d), we see the number of detected preparation erasures (orange markers) stays constant over the course of a 3 $\mu$s sweep; conversely, the number of detected decay erasures (green markers) grows over time, in good agreement with the measured Rydberg lifetime and erasure image infidelities (green solid line, Methods).

With the ability to distinguish these effects, we plot the total probability to form either of the AFM states, $P_\textrm{AFM} = P(\ket{grgr...gr}) + P(\ket{rgrg...rg})$ (Fig.~\ref{Fig3}e). At the conclusion of the sweep, we find $P_\textrm{AFM}=0.33^{+2}_{-2}$ without any erasure-excision (pink markers). By excising instances with preparation erasures, this fidelity is improved to $0.44^{+2}_{-2}$ (orange markers), and is then further improved to $0.49^{+2}_{-2}$ by additionally excising Rydberg decay erasures. The sharpness of the signal, exemplified by the derivative of $P_\textrm{AFM}$ with respect to the detuning, is similarly improved near the phase boundary (Fig.~\ref{Fig3}e, inset). We also observe that the gain in $P_\textrm{AFM}$ from erasure-excision increases with system size (Ext. Data Fig.~\ref{EFig_Nscaling}). 

We further explore how errors affect quantities reflecting higher-order statistics. To this end, we explore the probability distribution to find magnetic order of different magnitude by studying the AFM magnetization operator, defined as 
\begin{align}
\hat{M}=\hat{Z}_A/N_A-\hat{Z}_B/N_B,
\label{eq:afm}
\end{align}
where $\hat{Z}_{S}=\sum_{j\in S}\hat{Z}_j$ is the total magnetization operator in sub-lattice $S{=}A$ (odd sites) or $S{=}B$ (even sites) respectively, $N_{S}$ is the number of atoms in each sub-lattice, and $\hat{Z}_j=\ket{r}\bra{r}-\ket{g}\bra{g}$ is the local magnetization at site $j$. We plot the probability to find a specific eigenvalue, $M$, of $\hat{M}$ as a function of detuning (Fig.~\ref{Fig3}f). While the values of $M$ are initially tightly grouped around $M=0$ in the disordered phase, as the sweep progresses the probability distribution bifurcates, forming two separate dominant peaks in the $\mathbb{Z}_2$ phase, consistent with aforementioned two-fold spontaneous symmetry breaking across the quantum phase transition. We find that erasure-excision improves the sharpness of the distribution in both the disordered and $\mathbb{Z}_2$ phases (Fig.~\ref{Fig3}g). Near the phase transition, the distribution is close-to-flat, consistent with order appearing at all length scales. 

These results demonstrate improvements in fidelity for preparation of long-range-ordered ground states with erasure-excision in quantum simulation experiments, a first proof-of-principle for utilizing erasure conversion in NISQ-type applications.

\begin{figure}[t!]
	\centering
	\includegraphics[width=\columnwidth]{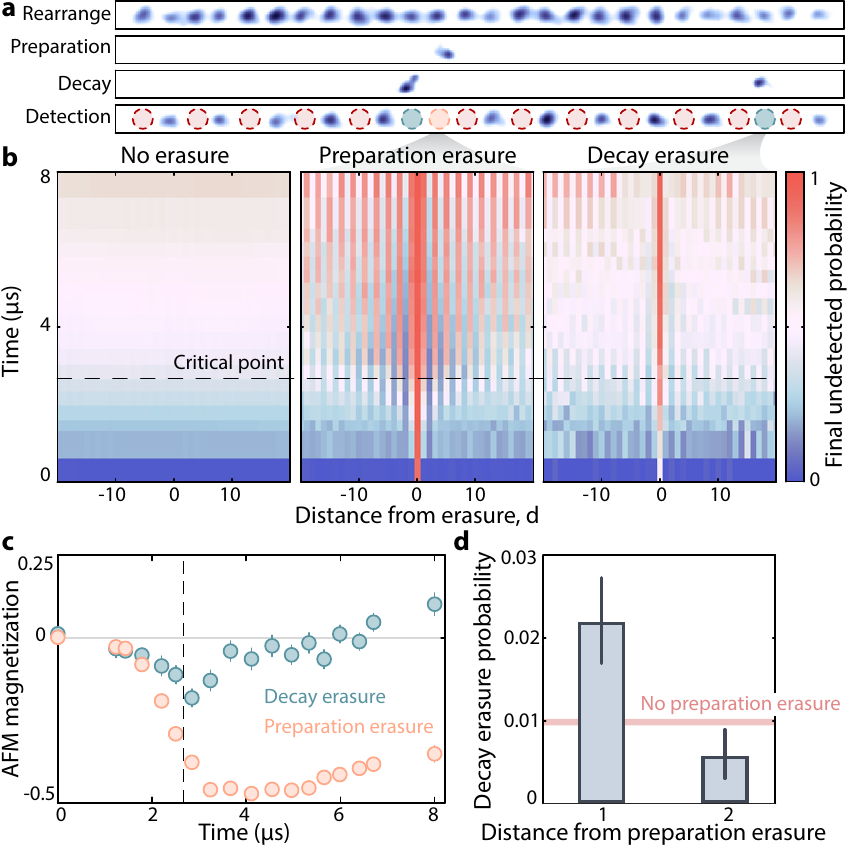}
	\caption{\textbf{Learning from erasure errors.} \textbf{a,} Post-processed~\cite{Bergschneider2018} single-shot atom fluorescence images (Methods). After arranging the array (top panel) we use the fast erasure images to learn how detected errors (middle panels) affect detection outcomes (lower panel). \textbf{b,} Conditional probability to detect no atom in the final image as a function of sweep progress and distance from a hypothetical erasure event. In the case of no erasures (left), the resulting profile is uniform. However, when conditioning on detecting a preparation erasure (middle), the error breaks the $\mathbb{Z}_2$ symmetry by establishing a single AFM order. In the case of conditioning on decay errors (right) the situation is more complex. \textbf{c,} AFM magnetization (see Eq.~\ref{eq:afm}) as a function of time. Preparation erasures (orange markers) lead to a growth of a single AFM order with Rydberg excitations predominantly on sub-lattice $A$ (defined as sites an odd distance from the erasure position). Decay erasures (green markers) follow a similar trend at early times by acting as effective preparation errors, but past the critical point (vertical dashed line) their behavior reverses: a decay erasure spontaneously breaks the two-fold symmetry, where the neighboring sites must have been in the ground state due to the $\mathbb{Z}_2$ structure, yielding Rydberg excitations on sub-lattice $B$. \textbf{d,} For the maximum sweep duration, the probability of detecting a Rydberg decay erasure (bars) is significantly increased (decreased) from the baseline level (red fill) at a distance of one (two) away from the site of a detected preparation erasure, induced by the altered Rydberg population on these sites due to the crystal formation in \textbf{b}, indicating direct detection of correlations between errors through erasure imaging.
 }
	\vspace{-0.5cm}
	\label{Fig5}
\end{figure}

\vspace{0.25cm}
\noindent \textbf{Learning from erasure errors}\\
Finally, we turn to studying a new tool enabled by our implementation of erasure conversion: exploring the effect of errors on experimental outcomes at a microscopic level and studying correlations between different error sources, which is enabled by having three separate images for a given experimental run (Fig.~\ref{Fig5}a). In particular, we consider the joint probability distribution, $\mathcal{P}(e^{(i)}_1,e^{(j)}_2,e^{(k)}_3)$, that atoms at sites $i$, $j$, and $k$ are detected respectively in the preparation erasure image ($e_1$), the decay erasure image ($e_2$) and the final state detection image ($e_3$). 

We again consider adiabatic sweeps into the $\mathbb{Z}_2$ phase as in Fig.~\ref{Fig3}, but now with a total duration of 8 $\mu$s. We first study $\mathcal{P}(e^{(j+d)}_3=0|e^{(j)}_1=0)$, equivalent to finding a Rydberg excitation on site $j+d$, conditioned on finding no preparation erasure on site $j$. We plot this quantity (Fig.~\ref{Fig5}b, left) as a function of both $d$ and the sweep duration. We explicitly average over choices of $j$ and find a signal essentially uniform in $d$.

However, if we instead consider $\mathcal{P}(e^{(j+d)}_3=0|e^{(j)}_1=1)$, the probability to find a Rydberg excitation on site $j+d$ conditioned on \textit{detecting} a preparation erasure on site $j$, markedly different behavior emerges (Fig.~\ref{Fig5}b, middle). For simplicity, we further post-select on instances where only a single erasure is detected across the entire array. At intermediate sweep times, we observe an AFM order forms around the preparation erasure error position. We interpret the error as breaking the atom chain into two shorter chains; excitations will naturally form at the system edges of these shorter chains in order to maximize the Rydberg density in the attractive regime in which we operate (Methods). This effectively pins the Rydberg density around the error, which then establishes a preferred AFM order further out into the array. Interestingly, the equivalent quantity for decay erasures, $\mathcal{P}(e^{(j+d)}_3=0|e^{(j)}_2=1)$, shows a more complex behavior.

To quantify this behavior more explicitly, we consider a variant of the AFM magnetization (Eq.~\ref{eq:afm}) conditioned on the erasure location, where sublattice $A$ ($B$) is now defined as being sites an odd (even) distance away from an erasure. In Fig.~\ref{Fig5}c we plot the mean AFM magnetization for both the preparation (orange circles) and decay erasure (green circles) cases. Preparation erasures develop a negative, single AFM order as they pin Rydberg excitations at odd distances away from the erasure. 

Decay erasures behave similarly before the critical point, as Rydberg decay acts effectively as a preparation error. However, past the critical point, this behavior changes: decay now acts as a measurement on the AFM superposition ground state, selecting one of these orders. In this case, assuming perfect $\mathbb{Z}_2$ states, the neighboring sites must have been in the ground state to detect a decay, meaning the AFM order is reversed from the preparation case. This leads to data that first dips to negative values and then grows to positive values past the phase transition (green markers in Fig.~\ref{Fig5}c).

We also study correlations between preparation errors and Rydberg decay. In particular, a preparation error forces atoms at odd intervals from the preparation erasure to have a higher probability to be in Rydberg states, meaning they should also be more likely to decay. As shown in Fig.~\ref{Fig5}d, we directly observe this effect at the end of the sweep by considering $\mathcal{P}(e^{(j+d)}_2=1|e^{(j)}_1=1)$, the probability to detect a decay erasure at a distance $d$ away from a preparation erasure. For $d=1$ ($d=2$), this probability is significantly increased (decreased) from the unconditional decay erasure probability, in line with the increased (decreased) Rydberg population on these sites, which shows that errors are correlated.

Before concluding, we note that erasure-excision for preparation errors using the first erasure image can be considered heralding the subsequent quantum simulation on the presence of atoms in tweezers in the correct initial state. For erasure-excision of Rydberg decay using the second erasure image, we interpret the post-selected results as coming from a non-jump trajectory in a Monte Carlo wavefunction approach~\cite{Carmichael1993}. 

\vspace{0.25cm}
\noindent \textbf{Discussion and Outlook}\\
Our results could have broad implications for quantum science and technology. First, our two-qubit entanglement fidelity values and associated error modelling imply that Rydberg arrays, which have already demonstrated scalability to hundreds of atoms~\cite{Scholl2021,Ebadi2021}, can be simultaneously equipped with high-fidelity two-qubit operations, a unique combination across all platforms. Besides our current demonstration of ${\sim}0.999$ SPAM corrected two-qubit fidelity, modeling implies that values of ${\sim}0.9997$ could be possible with laser noise improvements alone. Further, utilizing a cryogenic environment could freeze out blackbody decay to a large degree~\cite{Schymik2021}, with remaining decay detected as an erasure, leaving almost no intrinsic decoherence. In this context, we note very recent results for improved computational gate fidelities~\cite{Evered2023}.

Second, the demonstrated erasure conversion techniques could find wide-spread applications for both classical and quantum error correction. For classical correction, our techniques could be modified to correct for state-preparation errors via subsequent atom rearrangement~\cite{Endres2016,Barredo2016}, instead of just excising such events. Further, thermal excitations could be converted to erasures and subsequently removed by driving a blue sideband transition between $^1S_0$ and $^3P_0$ (Fig.~\ref{Fig1}a) prior to the fast image and subsequent atom-rearrangement~\cite{Endres2016, Barredo2016}, effectively realizing erasure-based atomic cooling.

For quantum-error correction, our techniques could be combined with a long-lived qubit which is dark to the fast image, e.g., realized with the $^3P_0$ nuclear qubit in neutral Sr~\cite{Barnes2022} and Yb~\cite{Jenkins2022, Ma2022}, or $S_{1/2}$ in Ca$^+$ and Ba$^+$ ions~\cite{Kang2023}. Similarly, schemes for implementing erasure conversion in superconducting circuits have been put forward~\cite{Kubica2022,Teoh2022}. Such techniques could lead to drastically reduced quantum error-correction thresholds~\cite{Wu2022,Sahay2023} for fault-tolerant quantum computing.

Third, our results also show clearly how NISQ applications~\cite{Preskill2018} can benefit from erasure conversion. Our demonstrated improvements for analog quantum simulation of ground-state physics could be extended to non-equilibrium dynamics, for example targeting regimes generating large entanglement entropies~\cite{Choi2023}, with the potential to reach a quantum advantage over classical simulations~\cite{Arute2019}. We note that while our implementation of erasure-excision slows down the effective sampling rate of the quantum device (Ext. Data Fig.~\ref{EFig_errthresh}), the classical cost can increase highly non-linearly with the resulting fidelity increase, and we hence expect a gain for such tasks. Further, we envision erasure-excision improving other tasks such as quantum optimization~\cite{Ebadi2022} and potentially quantum metrology~\cite{Pezze2018}.

Finally, insights into erasure-error correlations, as in Fig.~\ref{Fig5}, could be used to understand error processes in NISQ devices in unprecedented detail, in particular if erasure detection could be made time-resolved with respect to the many-body dynamics. This could also be used to realize post-measurement physics with erasure detection, such as measurement-induced phase transitions~\cite{Skinner2019,Li2018} and measurement-altered quantum criticality~\cite{Garratt2022}. 

\textit{Note---}During completion of this work we became aware of work performing erasure detection with ytterbium atoms~\cite{Ma2023}.

\begin{acknowledgements}
We acknowledge insightful discussions with, and feedback from, Hannes Pichler, Hannes Bernien, John Preskill, Jacob Covey, Chris Pattinson, Kevin Slagle, Hannah Manetsch, Jeff Thompson, Kon Leung, Elie Bataille, and Ivaylo Madjarov. We acknowledge support from the Institute for Quantum Information and Matter, an NSF Physics Frontiers Center (NSF Grant PHY-1733907), the DARPA ONISQ program (W911NF2010021), the NSF CAREER award (1753386), the AFOSR YIP (FA9550-19-1-0044), the NSF QLCI program (2016245), and the U.S. Department of Energy, Office of Science, National Quantum Information Science Research Centers, Quantum Systems Accelerator. PS acknowledges support from the IQIM postdoctoral fellowship. RBST acknowledges support from the Taiwan-Caltech Fellowship. RF acknowledges support from the Troesh postdoctoral fellowship.
\end{acknowledgements}

\section*{Data availability}
The data and codes that support the findings of this study are available from the corresponding author upon reasonable request.

\section*{Competing interests}
The authors declare no competing interests.

\section*{Author Contributions}
P.S., A.L.S., and M.E. conceived the idea and experiment. P.S., A.L.S, R.B.T., R.F. and J.C. performed the experiments, data analysis, and numerical simulations. P.S., A.L.S., R.B.T., R.F., and J.C. contributed to the experimental set-up. P.S., A.L.S. and M.E. wrote the manuscript with input from all authors. M.E. supervised this project.

\newpage
\clearpage

\clearpage

\section{Methods}

\subsection*{Fast imaging on the erasure detection subspace}
Here we describe how we perform the erasure imaging which allows us to detect site-resolved leakage errors~\cite{Bergschneider2018}. In order to both avoid any extra heating coming from the imaging beams and optimize imaging fidelity, we shine two identical counter-propagating beams with crossed $\pi$-polarization and Rabi frequencies $\Omega/2 \pi \simeq 40 \, \text{MHz}$ on the $^1S_0 \rightarrow {}^1P_1$ transition (see Ext. Data Fig.~\ref{EFig1}a). This minimizes the net force on an atom, and the crossed polarization avoids intensity interference patterns. 

We highlight the characteristic features of this imaging scheme experimentally. We show in Ext. Data Fig.~\ref{EFig1}b the survival probability of atoms in $^1S_0$ as a function of imaging time. After $4\, \mu \text{s}$, more than $80\%$ of the atoms are lost. However, the number of detected photons continues to increase: even though the kinetic energy of the atoms is too large to keep them trapped, their mean position remains centered on the tweezers. Importantly, for our implementation of erasure-excision, atom loss during the erasure image is inconsequential for our purposes as long as the initial presence of the atom is correctly identified, but in any case, other fast imaging schemes may alleviate this effect~\cite{Deist2022}. After ${\sim} 24 \, \mu \text{s}$, the atomic spread becomes too large and the number of detected photons plateaus. The obtained detection histogram is shown in Ext. Data Fig.~\ref{EFig1}c. We present the results both for empty (blue) and filled (red) tweezers, which we achieve by first imaging the atoms using usual, high survival imaging for initial detection in a $50\%$ loaded array, then perform the fast image. We obtain a typical detection fidelity of $0.980^{+1}_{-1}$ of true positives and true negatives, limited by the finite probability for atoms in $^1P_1$ to decay into $^1D_2$ (see Ext. Data Fig.~\ref{EFig1}a).

This imaging scheme is sufficiently fast to avoid perturbing atoms in $^3P_0$, as measured by losses from $^3P_0$ as a function of imaging time (Ext. Data Fig.~\ref{EFig1}d). We fit the data (circles) using a linear function (solid line), and obtain a loss of $0.0000046^{+12}_{-12}$ per image, consistent with the lifetime of the $^3P_0$ state~\cite{Shaw2023A} of ${\sim}5\, $s for the trap depth of $45 \, \mu \text{K}$ used during fast imaging.

As to the nature of the detected erasure errors for the Bell state generation, we find that preparation errors contribute the vast majority of erasure events compared to bright Rydberg decay, and excising them has a more significant impact on reducing infidelities. In particular, application of $\hat{U}$ lasts for only ${\sim}59\, $ns, which is significantly shorter than the independently measured bright state decay lifetime of $168^{+14}_{-14}\, \mu$s (Ext. Data Fig.~\ref{EFig_decay}). The error model described in Fig.~\ref{Fig2} suggests that excising such errors results in an infidelity reduction of only $1.2^{+3}_{-3}\times10^{-4}$ (Methods). Conversely, preparation errors account for ${\sim} 5\times10^{-2}$ infidelity per pair due to the long time between preparation in $\ket{g}$ and Rydberg excitation (Ext. Data Fig.~\ref{EFig_sequence}). Hence, the gains in fidelity from erasure-conversion mainly come from nearly eliminating all preparation errors, which has the added benefit of significantly reducing error bars on the SPAM corrected values. Still, SPAM corrected values might also benefit from the small gain in eliminating the effect of bright state decay, and from avoiding potential deleterious effects arising from higher atomic temperature in the repumper case. 

For erasure detection employed in the context of many-body quantum simulation, we adjust the binarization threshold for atom detection to raise the false positive imaging fidelity to 0.9975, while the false negative imaging fidelity is lowered to ${\sim}0.6$ (Fig.~\ref{Fig3}d); this is done as a conservative measure to prioritize maximizing the number of usable shots while potentially forgoing some fidelity gains (Ext. Data Fig.~\ref{EFig_errthresh}).

We note that the scheme we show here is not yet fundamentally limited, and there are a number of technical improvements which could be made. First, the camera we use (Andor iXon Ultra 888) has a quantum efficiency of $\sim80\%$, which has been improved in some recent models, such as qCMOS devices. Further, we presently image atoms only from one direction, when in principle photons could be collected from both objectives~\cite{Graham2022}. This would improve our estimated total collection efficiency of $\sim4\%$ by a factor of 2, leading to faster imaging times with higher fidelity (as more photons could be collected before that atoms were ejected from the trap). Furthermore, the fidelity may be substantially improved by actively repumping the $^1D_2$ state back into the imaging manifold so as to not effectively lose any atoms via this pathway.

\subsection*{Details of Rydberg excitation}
Our Rydberg excitation scheme has been described in depth previously~\cite{Madjarov2020}. Prior to the Rydberg excitation, atoms are initialized from the absolute ground state 5s$^2$ $^1S_0$ to the metastable state 5s5p $^3P_0$ (698.4 nm) through coherent drive. Subsequently, tweezer trap depths are reduced by a factor of 10 to extend the metastable state lifetime.

For Rydberg excitation and detection, we extinguish the traps, drive to the Rydberg state (5s61s $^3S_1, m_J=0$, 317 nm), and finally perform auto-ionization of the Rydberg atoms~\cite{Madjarov2020}. Auto-ionization has a characteristic timescale of ${\sim}5$ ns, but we perform the operation for 500 ns to ensure total ionization. We report a more accurate measurement of the auto-ionization wavelength as ${\sim}407.89$ nm. In the final detection step, atoms in $^3P_0$ are readout via our normal imaging scheme~\cite{Madjarov2020,Covey2019A}.

Atoms can decay from $^3P_0$ between state preparation and Rydberg excitation, which is 60 ms to allow time for magnetic fields to settle. In previous work~\cite{Madjarov2020}, we supplemented coherent preparation with incoherent pumping to $^3P_0$ immediately prior to Rydberg operations. However, during the repumping process, atoms can be lost due to repeated recoil events at low trap depth, which is not detected by the erasure image, and thus can lower the bare fidelity. Even with SPAM-correction of this effect, we expect the fidelity with repumping to be slightly inferior due to an increased atomic temperature for pumped atoms.

\subsection*{Rydberg Hamiltonian}
The Hamiltonian describing an array of Rydberg atoms is well approximated by \begin{align}
\hat{H}/\hbar=\frac{\Omega}{2}\sum_i \hat{X}_i -\Delta\sum_i \hat{n}_i + \frac{C_6}{a^6} \sum_{i>j} \frac{\hat{n}_i \hat{n}_j}{|i-j|^6} \label{eq:RydbergHam}
\end{align}
which describes a set of interacting two-level systems, labeled by site indices $i$ and $j$, driven by a laser with Rabi frequency $\Omega$ and detuning $\Delta$. The interaction strength is determined by the $C_6$ coefficient and the lattice spacing $a$. Operators are $\hat{X}_i =\ket{r}_i\bra{g}_i + \ket{g}_i \bra{r}_i$ and $\hat{n}_i = \ket{r}_i\bra{r}_i$, where $\ket{g}_i$ and $\ket{r}_i$ denote the metastable ground and Rydberg states at site $i$, respectively. 

For the case of measuring two-qubit Bell state fidelities we set $\Omega/2\pi=6.2$ MHz. Interaction strengths in Fig.~\ref{Fig2}a are directly measured at inter-atomic separations of 4 and 5 $\mu$m, and extrapolated via the predicted $1/r^6$ scaling to the level at 2.5 $\mu$m. Mean atomic distances are calibrated via a laser-derived ruler based on shifting atoms in coherent superposition states~\cite{Shaw2023B}. We calibrate $C_6/2\pi =230(25)$ GHz~$\mu m^6$ using maximum-likelihood-estimation (and associated uncertainty) from resonant quench dynamics~\cite{Choi2023}, which additionally calibrates a systematic offset in our global detuning.

For performing many-body quasi-adiabatic sweeps, the detuning is swept symmetrically in a tangent profile from +30 MHz to -30 MHz, while the Rabi frequency is smoothly turned on and off with a maximum value of $\Omega/2\pi=5.6$ MHz. For an initially positive detuning, the $\ket{r}$ state is energetically favorable, making the all ground initial state, $\ket{gg...gg}$, the highest energy eigenstate of the blockaded energy sector, where no neighboring Rydberg excitations are allowed. For negative detunings, where $\ket{g}$ is energetically favorable, the highest energy state uniquely becomes the symmetric AFM state $(\ket{grgr...gr}+\ket{rgrg...rg})/\sqrt{2}$ in the deeply ordered limit. Thus, considering only the blockaded energy sector, sweeping the detuning from positive to negative detuning (thus remaining in the highest energy eigenstate) is equivalent to the ground state physics of an effective Hamiltonian with attractive Rydberg interaction and inverted sign of the detuning. This equivalence allows us to operate in the effectively attractive regime of the blockaded phase diagram of Ref.~\cite{Fendley2004}. For our Hamiltonian parameters, we use exact diagonalization numerics to identify the infinite-size critical detuning using a scaling collapse near the finite-system size minimum energy gap~\cite{Samajdar2018}.

\subsection*{Error modeling}
Our error model has been described previously~\cite{Madjarov2020,Choi2023}. We perform Monte Carlo wavefunction based simulations~\cite{Leseleuc2018}, accounting for a variety of noise sources including: time-dependent laser intensity noise, time-dependent laser frequency noise, sampling of the beam intensity from the atomic thermal spread, Doppler noise, variations of the interaction strength from thermal spread, beam pointing stability, and others. All of the parameters which enter the error model are independently calibrated via selective measurements directly on an atomic signal if possible, as shown in Ext. Data Table~\ref{EFig:table}. Parameters are not fine-tuned to match the measured Bell state fidelity, and the model equally well describes results from many-body quench experiments~\cite{Choi2023}.

\subsection*{Extraction of the Bell state fidelity}
In order to extract the Bell state fidelities quoted in the main text, we use a lower bound method~\cite{Madjarov2020} which relies on measuring the populations in the four possible states $P_{gr}$, $P_{rg}$, $P_{gg}$ and $P_{rr}$ during a Rabi oscillation between $\ket{gg}$ and $\ket{\Psi^+}$. The lower bound on Bell state fidelity is given by:
\begin{align}
F_\text{Bell} \geq \frac{P_{gr+rg}^\pi}{2} +\sqrt{ \frac{\sum_i (P_{i}^{2\pi})^2-1}{2} + P_{gr}^\pi P_{rg}^\pi},
\label{eq:Bell}
\end{align} 
where $P_i^{2\pi}$ are the measured probabilities for the four states at $2\pi$, and $P_{gr+rg}^\pi$ is the probability $P_{gr} + P_{rg}$ measured at $\pi$. In order to measure these probabilities with high accuracy, we concentrate our data-taking around the $\pi$ and $2\pi$ times (Ext. Data Fig.~\ref{EFig2}a), and fit the obtained values using quadratic functions $f(t) = p_0 + p_1(t-p_2)^2$. We first detail the fitting method, then how we obtain the four probabilities, and finally the extraction of the Bell state fidelity from these.

\subsubsection*{Fitting method}

We perform a fit which takes into account the underlying Beta distribution of the data, and prevents systematic errors arising from assuming a Gaussian distribution of the data. The aim of the fit is to obtain the three-dimensional probability density function $Q(p_0,p_1,p_2)$ of $f$, using each experimental data point $i$ defined by its probability density function $\mathcal{P}_i(x)$, where $x$ is a probability. In order to obtain a particular value of $Q(\tilde{p}_0,\tilde{p}_1,\tilde{p}_2)$, we look at the corresponding probability density function value $\mathcal{P}_i(f(t_i))$ for each data point $i$, where $f(t_i) = \tilde{p}_0 + \tilde{p}_1(t_i-\tilde{p}_2)^2$, and assign the product of each $\mathcal{P}_i(f(t_i))$ to the fit likelihood function: 
\begin{align}
Q(\tilde{p}_0,\tilde{p}_1,\tilde{p}_2) = \prod_{i} \mathcal{P}_i(f(t_i)). 
\end{align}
We repeat this for various $[\tilde{p}_0,\tilde{p}_1,\tilde{p}_2]$. 

The result of such fitting method is shown in Ext. Data Fig~\ref{EFig2}b (black line), where we present $f(t) = p_0 + p_1(t-p_2)^2$ for $[p_0,p_1,p_2]$ corresponding to the maximum value of $Q(p_0,p_1,p_2)$. We emphasize that this results in a lower peak value than a standard fitting procedure which assumes underlying Gaussian distributions of experimentally measured probabilities (red line). Choosing this lower peak value eventually will provide a more conservative, but more accurate value for the Bell state fidelity lower bound than the na\"ive Gaussian approach.

\subsubsection*{Obtaining the four probability distributions}

Our method to obtain the probability density functions of the four probabilities at $\pi$ and $2\pi$ times ensures both that the sum of the four probabilities always equals one, and that their mutual correlations are preserved. We first extract the Beta distribution of $P_{rr}$ by gathering all the data around the $\pi$ and $2\pi$ times (Ext. Data Fig.~\ref{EFig2}c). In particular, the mode of the obtained Beta distribution at $\pi$ is $P_{rr} \simeq 0.0005$. The distribution of $P_{gr+rg}$ and $P_{gg}$ are obtained by fitting the data in the following way.
We perform a joint fit on $P_{gr+rg}$ using a fit function $f_1(t)$, and on $P_{gg}$ using a fit function $f_2(t)$. The fit functions are expressed as: 
\begin{align}
f_1(t) = \ & p_0 + p_1(t-p_2)^2,  \\
f_2(t) = \ & 1-p_0-P_{rr} - p_1(t-p_2)^2,
\end{align} 
which ensures that the sum of the four probabilities is always equal to 1.
We then calculate the joint probability density function $Q_{1,2}(p_0,p_1,p_2)$ of both $f_1$ and $f_2$ using the method described above. In particular:
\begin{align}
Q_{1,2}(\tilde{p}_0,\tilde{p}_1,\tilde{p}_2) = \prod_{i} \mathcal{P}_i^{gr+rg}(f_1(t_i)) \prod_{i} \mathcal{P}_i^{gg}(f_2(t_i)),
\end{align}
where $\mathcal{P}_i^{gr+rg}$ ($\mathcal{P}_i^{gg}$) is the probability density function associated with $P_{gr+rg}$ ($P_{gg}$) for the $i$-th experimental data point. In particular, we impose that $p_0\leq 1-P_{rr}$ to avoid negative probabilities. We show the resulting $Q_{1,2}(p_0,p_1,p_2)$ in Ext. Data Fig.~\ref{EFig2}d as two-dimensional maps along $(p_0,p_1)$ and $(p_0,p_2)$.

We then obtain the one dimensional probability density function for $p_0$ by integrating over $p_1$ and $p_2$ (see Ext. Data Fig.~\ref{EFig2}d). This provides the fitted probability density function of $P_{gr+rg}$, and hence $P_{gg} = 1-P_{rr}-P_{gr}-P_{rg}$ at $\pi$ time. We repeat this process for various values of $P_{rr}$, for both $\pi$ and $2\pi$ times.

At the end of this process, we obtain different probability density functions for each $P_{rr}$ value. The asymmetry between $P_{gr}$ and $P_{rg}$ is obtained by taking the mean of $P_{gr} - P_{rg}$ at $\pi$ and $2\pi$ times. We assume the underlying distribution to be Gaussian, as $P_{gr}-P_{rg}$ is centered on 0, and can be positive or negative with equal probability.

\subsubsection*{Bell state fidelity}

Now that we have the probability density function for all four probabilities at $\pi$ and $2\pi$ times, we move on to the Bell state fidelity extraction. For both $\pi$ and $2\pi$, we perform a Monte-Carlo sampling of the Beta distribution of $P_{rr}$ which then leads to a joint probability density function for $P_{gr+rg}$ and $P_{gg}$. We then sample from this, and use Eq.~\ref{eq:Bell} to obtain a value for the Bell state fidelity lower bound. We repeat this process 1 million times, and fit the obtained results using a Beta distribution (see Ext. Data Fig.~\ref{EFig2}e). We observe an excellent agreement between the fit and the data, from which we obtain $F_\text{Bell} \geq 0.9962^{+10}_{-13}$, where the quoted value is the mode of the distribution, and the error bars represent $68\%$ confidence interval.

We use the same method to obtain the measurement-corrected Bell fidelity, and the state preparation and measurement (SPAM) corrected one. After drawing the probabilities from the probability density functions, we infer the SPAM-corrected probabilities from our known errors, described in detail previously~\cite{Madjarov2020}. We use the values reported in Ext. Data Table~\ref{EFig:table}. During this process, there is a finite chance that the sum of probabilities does not sum up to one. This comes from the fact that the probability density functions and the SPAM correction are uncorrelated, an issue which is avoided for raw Bell fidelity extraction thanks to the correlated fit procedure described above. We employ a form of rejection sampling to alleviate this issue by restarting the whole process in the case of such event. We perform this 1 million times, and fit the obtained results using a Beta distribution (see Ext. Data Fig.~\ref{EFig2}f). We observe an excellent agreement between the fit and the data, from which we obtain a SPAM-corrected fidelity $F_\text{Bell} \geq 0.9985^{+7}_{-12}$, where the quoted value is the mode of the distribution, and the error bars represent $68\%$ confidence interval.

\subsection*{Interaction limitation for Bell fidelity}
We estimate the theoretically expected Bell state fidelity using perturbation analysis. Specifically, the resonant blockaded Rabi oscillation for an interacting atom pair is described by the following Hamiltonian
\begin{align}
    \hat{H}/\hbar = \frac{\Omega}{2} (\hat{X}_1 + \hat{X}_2) + V \hat{n}_1 \hat{n}_2,
\end{align}
where $V = C_6/r^6$ is the distance-dependent, interaction strength between two atoms separated at distance $r$ (see Eq.~\ref{eq:RydbergHam}). Since the two-atom initial ground state, $\ket{\psi(0)} = \ket{gg}$, has even parity under the left-right reflection symmetry, the Rabi oscillation dynamics can be effectively solved in an even-parity subspace with three basis states of $\ket{gg}, \ket{rr}$, and $\ket{\Psi^+} = \frac{1}{\sqrt{2}}(\ket{gr}+\ket{rg})$. In the Rydberg-blockaded regime where $V \gg \Omega$, we can perform perturbation analysis with the perturbation parameter $\eta = \Omega/\sqrt{2}V$ and find that the energy eigenvectors of the subspace are approximated as 
\begin{align*}
    \ket{E_1} &\approx \frac{(1-\frac{\eta}{4}-\frac{\eta^2}{32})\ket{gg} + (-1-\frac{\eta}{4}+\frac{17\eta^2}{32})\ket{\Psi^+} + (\eta - \frac{3\eta^2}{4})\ket{rr}}{\sqrt{2}} \\
    \ket{E_2} &\approx \frac{(-1-\frac{\eta}{4}+\frac{\eta^2}{32})\ket{gg} + (-1+\frac{\eta}{4}+\frac{17\eta^2}{32})\ket{\Psi^+} + (\eta + \frac{3\eta^2}{4}) \ket{rr}}{\sqrt{2}} \\
    \ket{E_3} &\approx \eta^2 \ket{gg} + \eta \ket{\Psi^+} + \ket{rr}
\end{align*}
with their corresponding energy eigenvalues of $E_1 \approx V(-\eta - \eta^2/2), E_2 \approx V(\eta - \eta^2/2)$, and $E_3 \approx V(1+\eta^2/2)$, respectively. Rewriting the initial state using the perturbed eigenbasis, we solve
\begin{align}
    F_\text{Bell} = \max_{t} |\langle \Psi^+ | e^{-i\hat{H}t} |\psi(0)\rangle|^2
\end{align}
to obtain the analytical expression of the maximum achievable Bell state fidelity, $F_\text{Bell}$, at a given perturbation strength $\eta$. Keeping the solution up to the second order of $\eta$, we find
\begin{align}
    F_\text{Bell} = 1 - \frac{5}{4} \eta^2 = 1 - \frac{5}{8}\left(\frac{\Omega}{V}\right)^2
\end{align}
obtained at $t = \pi/\sqrt{2}\Omega$.

\subsection*{Statistics reduction due to erasure-excision}
Our demonstration of erasure-excision explicitly discards some experimental realizations (Ext. Data Fig.~\ref{EFig_Nscaling}), which can be seen as a downside of the method. However, this is a controllable trade-off: by adjusting the threshold for detecting an erasure error, we can balance gains in fidelity versus losses in experimental statistics (as shown in Ext. Data Fig.~\ref{EFig_errthresh}) for whatever particular task is of interest. In general, the optimum likely always includes some amount of erasure-excision, as it is usually better to remove erroneous data than keeping them.
\clearpage
\newpage
\FloatBarrier
\setcounter{figure}{0}
\captionsetup[figure]{labelfont={bf},name={Ext. Data Fig.},labelsep=bar,justification=raggedright,font=small}
\section*{Extended Data Figures}

\begin{figure*}[htb!]
	\centering
	\includegraphics[width=\textwidth]{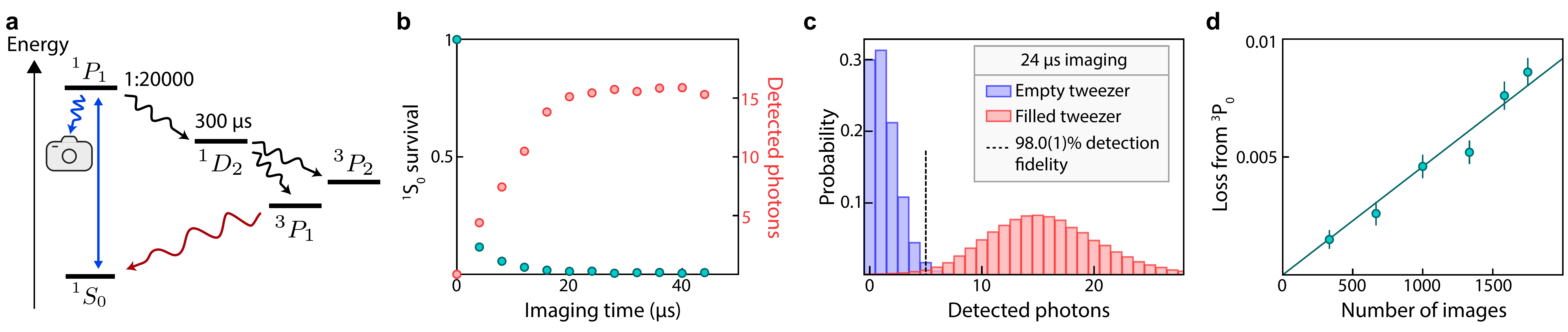}
	\caption{\textbf{Fast imaging on the erasure detection subspace.} \textbf{a,} Sketch of the involved energy levels. We detect atoms in $^1S_0$ by strongly driving the $^1S_0 \leftrightarrow {}^1P_1$ transition. \textbf{b,} Survival of atoms in  $^1S_0$ (green) and number of detected photons (red) as a function of the imaging time. We observe an increase of detected photons whereas the atoms are already lost: even though the kinetic energy of the atoms is too large to keep them trapped, their mean position remains centered on the tweezers thanks to the use of two counter-propagating beams with equal power. After ${\sim} 24 \, \mu \text{s}$, the atomic spread becomes too large to measure a significant increase in detected photons. \textbf{c,} Typical histograms of the number of detected photons for $24 \, \mu \text{s}$ imaging. Using a slow, high-fidelity image prior to the fast image, we can detect if a tweezer is empty (blue) or filled (red). The typical detection fidelity which corresponds to equal error probability in detecting absence or presence of an atom is $98.0(1)\%$. \textbf{d,} Losses from $^3P_0$ as a function of time, expressed in number of fast images. The survival probability of an atom in $^3P_0$ is $99.99954(12)\%$ for one image, consistent with its 5 second lifetime.
 }
	\label{EFig1}
\end{figure*}

\begin{figure*}[t!]
	\centering
	\includegraphics[width=\textwidth]{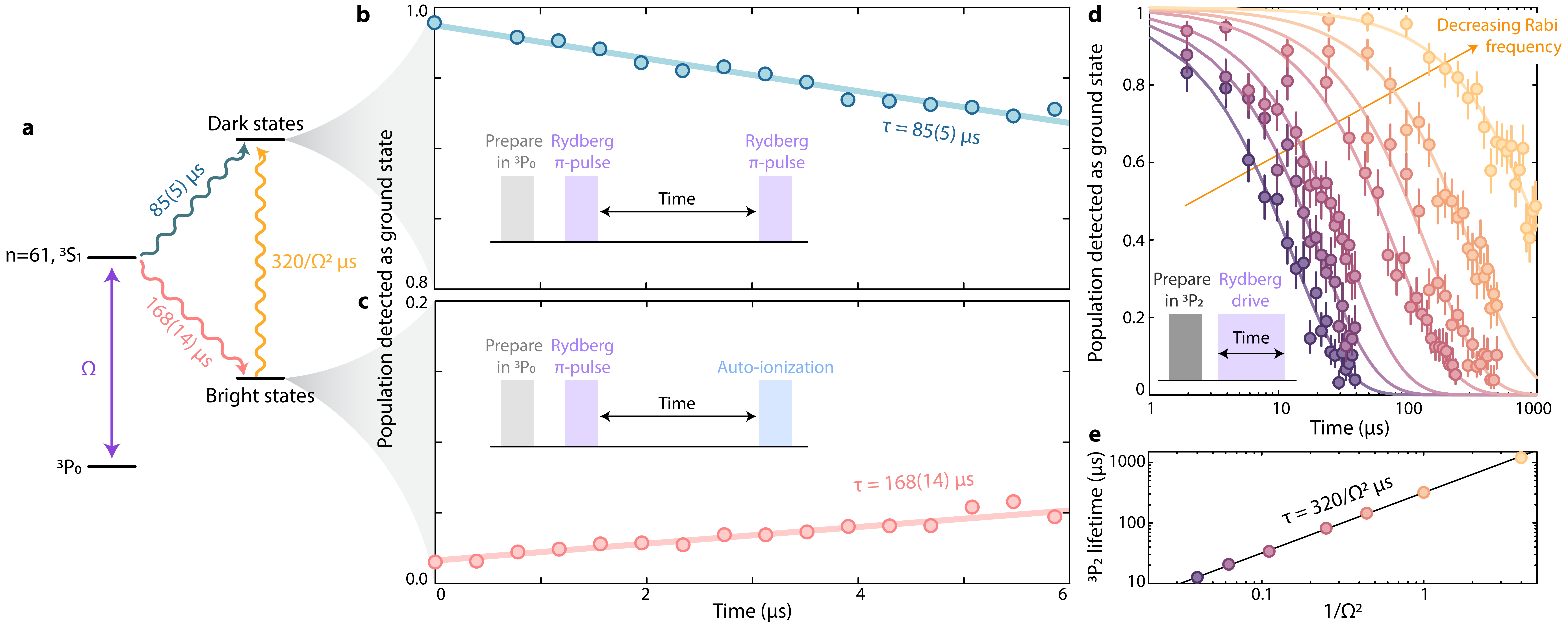}
	\caption{\textbf{Rydberg lifetime. a,} Pseudo-level diagram of the Rydberg dynamics and associated decay channels. During Rydberg evolution with Rabi frequency $\Omega$, the Rydberg atom can decay into either a set of states which is `bright' to the imaging process (including both the erasure images and the final detection images), e.g states like 5s5p ${}^3P_2$, or into states which are `dark' to the imaging process, e.g. nearby Rydberg states. A small percentage of decays into `bright' states can go directly into ${}^3P_0$ where they can be re-excited by the Rydberg driving; note that such decays are dark in the erasure image, but bright in the final detection image. \textbf{b,} Measurement of the dark state decay lifetime, measured by performing a $\pi$-pulse on the Rydberg transition, waiting a variable amount of time, and then returning atoms to the ground state (inset). \textbf{c,} Measurement of the bright state decay lifetime, measured by performing a Rydberg $\pi$-pulse, waiting, and then performing an auto-ionization pulse to destroy any remaining Rydberg or dark state excitations. \textbf{d,} We prepare atoms into ${}^3P_2$ (a bright state), and then perform continuous Rydberg driving. Atoms are lost from the trap at a rate which increases with increasing Rabi frequency. \textbf{e,} The lifetime of atoms in ${}^3P_2$ scales inversely with the square of the Rabi frequency (i.e. scales inversely with the intensity of the Rydberg beam). We attribute this to a photo-ionization process which can convert bright state decay into dark state decay through prolonged Rydberg excitation, as shown in \textbf{a}. Markers in \textbf{b-d} are experimental data where error bars are often smaller than the marker sizes, and solid lines represent exponential fits.}
	\label{EFig_decay}
\end{figure*}

\begin{figure*}[t!]
	\centering
	\includegraphics[width=\textwidth]{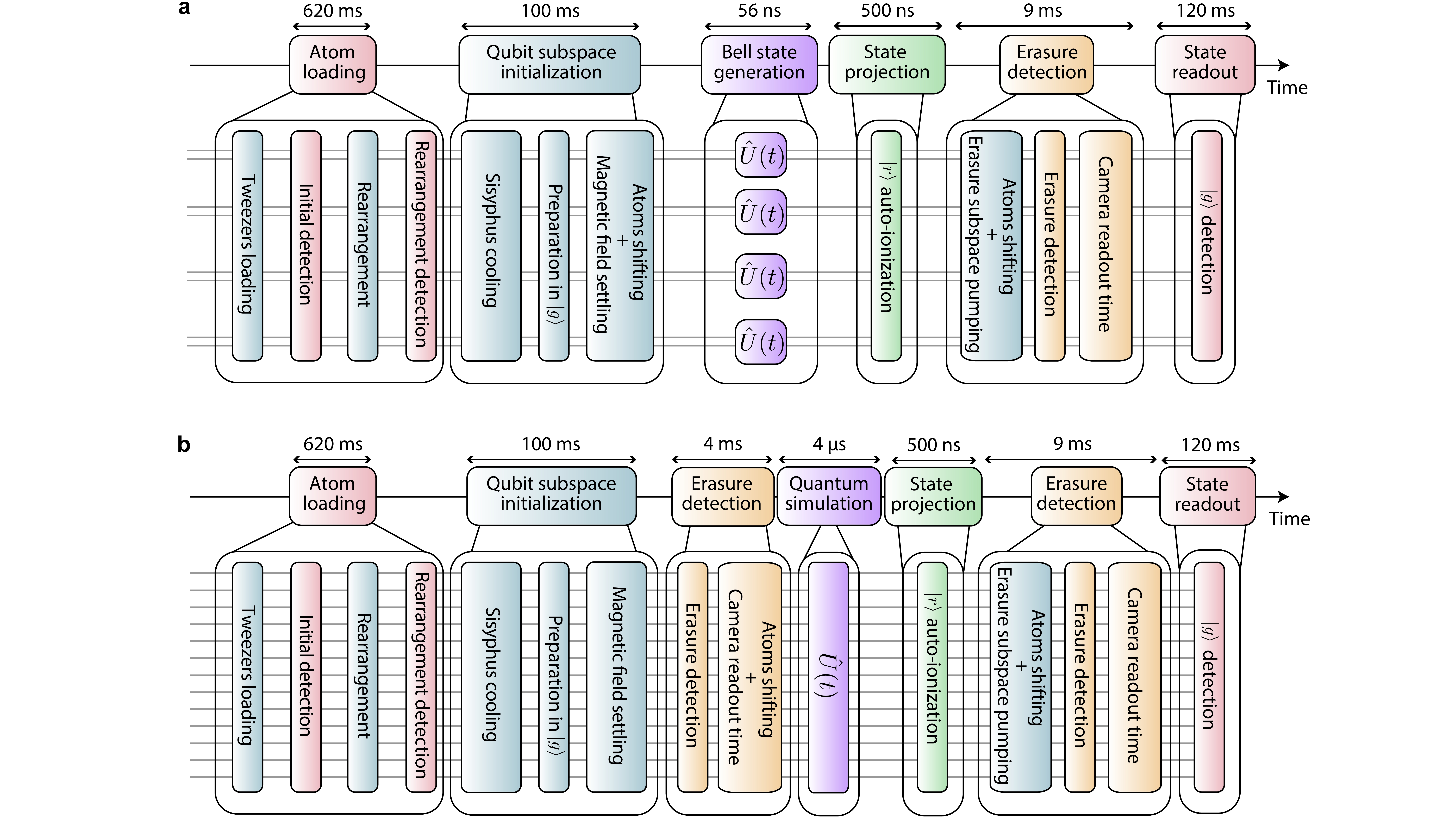}
	\caption{\textbf{Experimental sequence.} Sketch of the experimental sequence including the erasure detection for \textbf{a} the Bell state generation experiment, and \textbf{b} the many-body experiment. Both experiments have the same global architecture: we start by loading the atoms into the desired geometry, then initialize the atoms in $\ket{g}$, perform the Bell state generation or quantum simulation, and finally read out by auto-ionizing atoms in $\ket{r}$ and imaging atoms in $\ket{g}$. The main difference between both experiments concerns the erasure detection. In \textbf{a}, we utilize a single erasure detection, placed after auto-ionizing atoms in $\ket{r}$. In \textbf{b}, we perform two erasure images: one before applying $\hat{U}(t)$, and one after auto-ionization.}
	\label{EFig_sequence}
\end{figure*}

\begin{figure*}[t!]
	\centering
	\includegraphics[width=\textwidth]{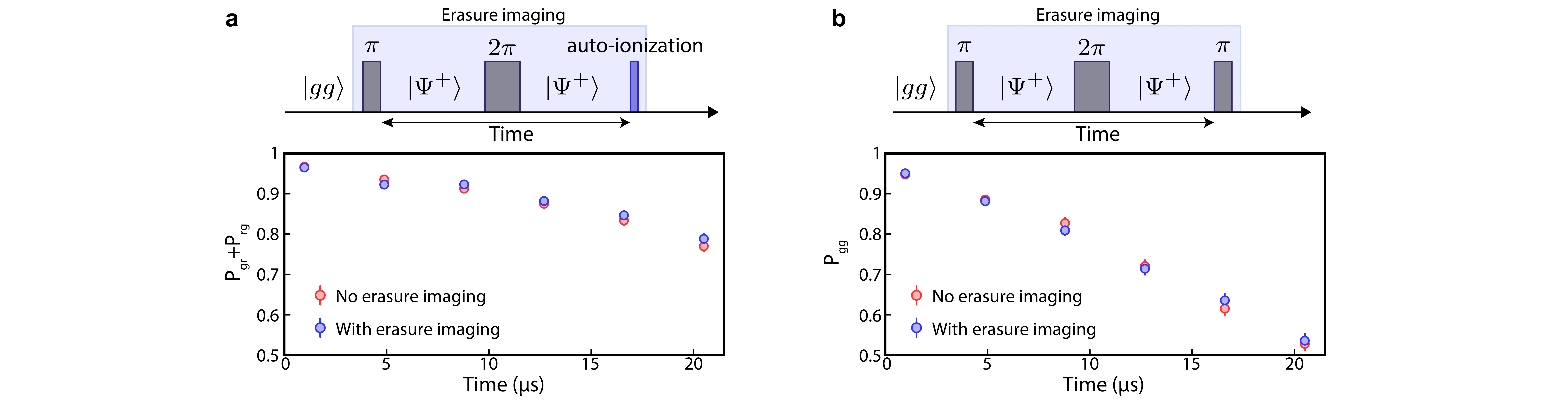}
	\caption{\textbf{Coherence preservation during erasure detection. a,b,} We prepare the Bell state $\ket{\Psi^+}$, and measure the relevant populations for Bell fidelity extraction $P_{gr}+P_{rg}$ and $P_{gg}$ (after an extra $\pi$ pulse) as a function of holding time. We perform a $2\pi$ pulse in the middle of the holding time to get rid of dephasing effect due to e.g. Doppler effect. We present the results with (blue) and without (red) performing the erasure imaging during the holding time. We observe no significant difference between the two conditions, which suggests that the erasure detection imaging light, in principle, does not destroy the coherence of the Bell state.}
	\label{EFig_coherence}
\end{figure*}

\begin{figure*}[t!]
	\centering
	\includegraphics[width=\textwidth]{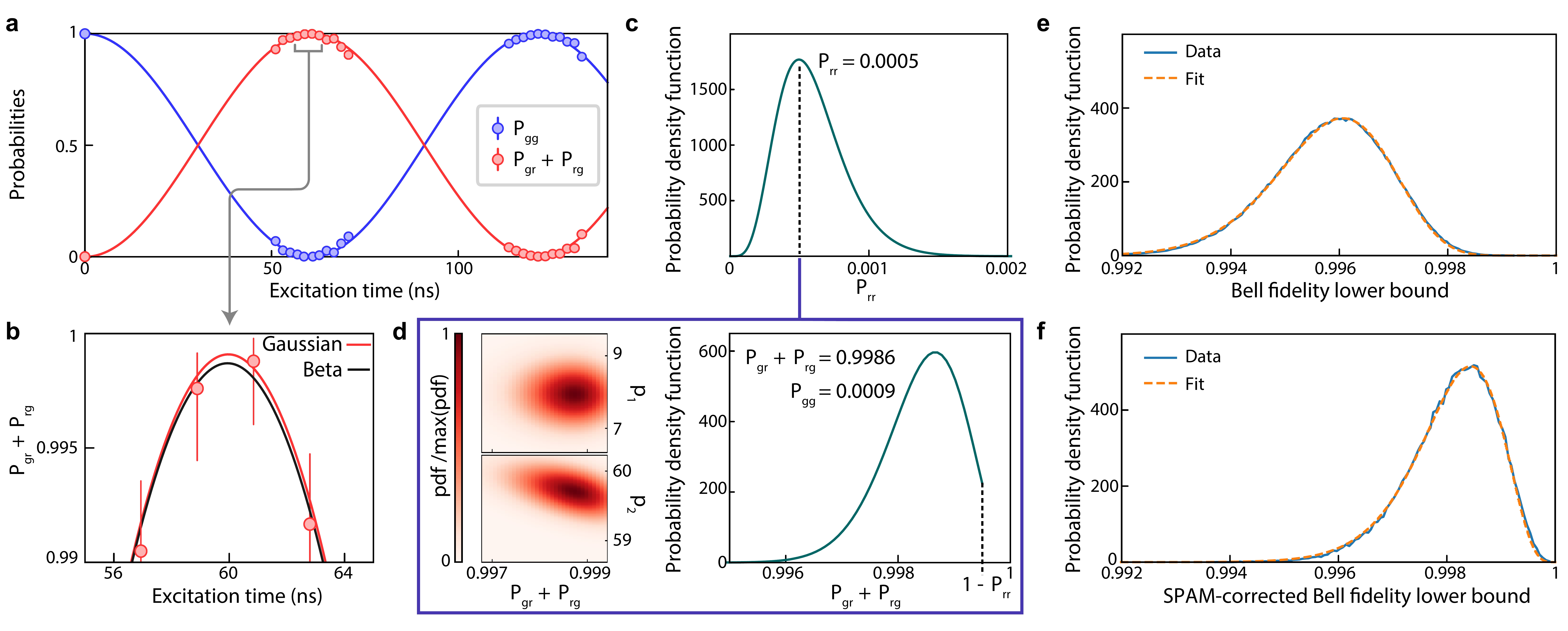}
	\caption{\textbf{Bell state fidelity measurement from blockaded Rabi oscillations.} \textbf{a} Probabilities of measuring both atoms in the ground state $P_\text{gg}$ (blue markers) and a single atom in the ground state $P_\text{gr} + P_\text{rg}$ (red markers) in the blockaded regime as a function of excitation time. Solid lines are guides to the eye. \textbf{b,} Zoom over the $\pi$ time, where we prepare the Bell state $\ket{\Psi^+}$ (see main text). We use a quadratic fit function of the form $f(x) = p_0+p_1(x-p_2)^2$ to extract the population values at $\pi$ and $2\pi$ times. We show the fit results (i) assuming the experimental data have a Gaussian uncertainty (red line), and (ii) using their true Beta distribution (black line). \textbf{c,d} Fitting method used to obtain the probability distributions of $P_\text{rr}$, $P_\text{gg}$, and $P_\text{gr} + P_\text{rg}$ at $\pi$ and $2\pi$ times. We first experimentally obtain the Beta distribution of the probability $P_\text{rr}$ to observe both atoms in the Rydberg state. We then perform a joint fit on $P_\text{gr} + P_\text{rg}$ and $P_\text{gg}$ with the same $p_1$ and $p_2$ fit coefficients for both. We fix the value of $P_\text{rr}$, and condition the joint fit such that the sum of all probabilities is always equals to one. We repeat this process for various values of $P_\text{rr}$. The results shown here are for $P_\text{rr}=0.0005$, which is the mode of the obtained Beta distribution for $P_\text{rr}$. The fitting method uses the true, experimentally measured Beta distribution of each data point. We obtain corresponding probability density functions for each $P_\text{rr}$. We perform this method independently for both $\pi$ and $2\pi$ times. \textbf{e,} Resulting Bell state fidelity lower bound using the probability density functions of $P_\text{rr}$ and $P_\text{gr} + P_\text{rg}$. We start by randomly drawing from the $P_\text{rr}$ distribution, then assign the corresponding probability density function of $P_\text{gr} + P_\text{rg}$, and draw a value from it. The asymmetry between $P_\text{gr}$ and $P_\text{rg}$ is obtained by averaging over each experimental data point, and is assumed to be Gaussian. We repeat this process 1 million times for both $\pi$ and $2\pi$ times. We obtain the corresponding probability density function (blue line), which we fit using a Beta distribution (orange dashed line). \textbf{f,} SPAM-corrected Bell state fidelity lower bound distribution, obtained by correcting the probabilities after randomly drawing them from their respective probability density functions.
 }
	\label{EFig2}
\end{figure*}

\begin{figure*}
	\centering
	\includegraphics[width=\textwidth]{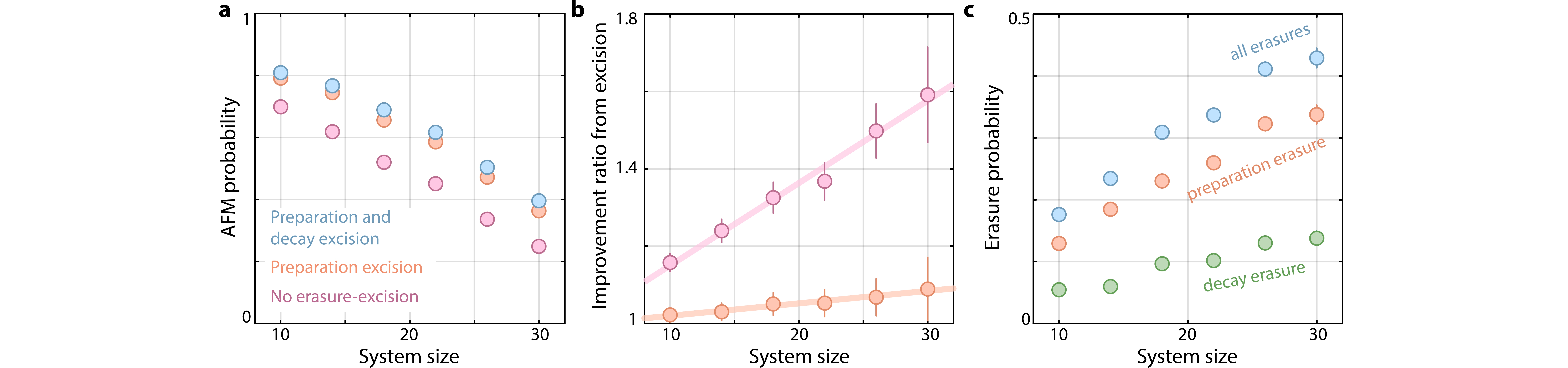}
	\caption{\textbf{System size scaling of sweep fidelity.} \textbf{a,} Total probability for forming either of the $\mathbb{Z}_2$ states following the quasi-adiabatic sweeps presented in Fig.~\ref{Fig3} of the main text, here presented for the final sweep time as a function of system size for full erasure-excision (blue markers), preparation erasure-excision (orange markers), and the baseline (pink markers) data. \textbf{b,} The ratio gain from using full erasure-excision grows as a function of system size, both with respect to the baseline values (pink markers), and to the case of only excising preparation erasures (orange markers). Solid lines are linear guides to the eyes. \textbf{c,} For a fixed detection threshold of 5 photons, the number of erasure errors also increases as a function of system size.
 }
	\label{EFig_Nscaling}
\end{figure*}

\begin{figure}[t!]
	\centering
	\includegraphics[width=\columnwidth]{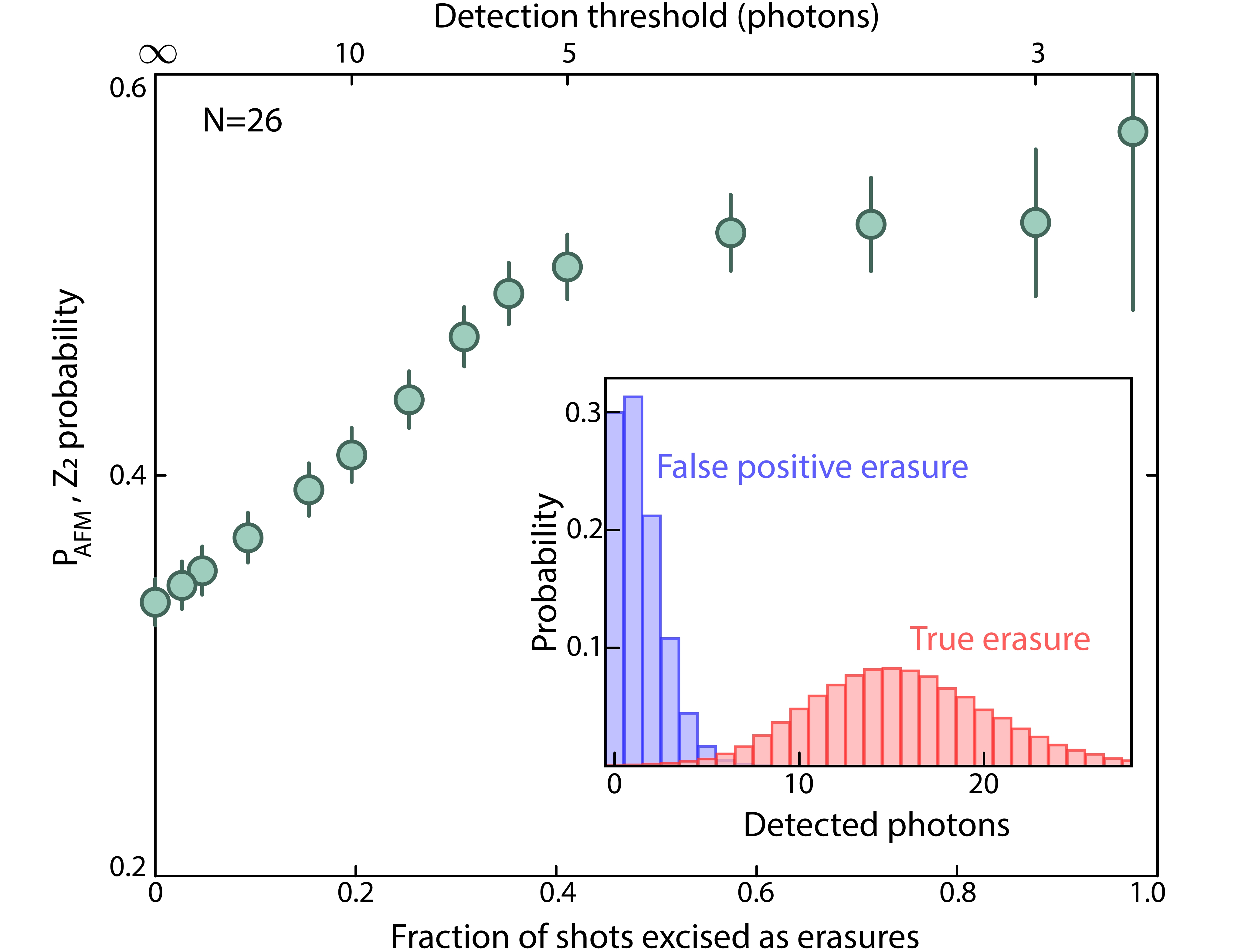}
	\caption{\textbf{Controlling fidelity gains from erasure-excision.} Erasure-excision explicitly trades improved experimental fidelity for a reduced number of experimental repetitions. However, this process is controllable by adjusting the threshold used for detecting atoms during erasure images. Changing the threshold essentially changes the false positive and false negative rate for detecting erasures correctly (inset). We plot the total AFM probability (green markers) after a sweep as in Fig.~\ref{Fig3} of the main text, and vary the detection threshold used for identifying erasures. For too high a threshold, many erasure events go unnoticed, and so erroneous outcomes become relatively more prevalent, reducing the overall fidelity. As the detection threshold is lowered, more true erasures, where an atom is actually present in the erasure image, are correctly detected, which improves the fidelity. However, lowering the threshold too far (in our case past ${\sim}5$ photons) increases the likelihood of seeing false positive erasures; excising data based on these events discards experimental statistics with relatively little gain in fidelity. In the main text, we select a detection threshold of 5 photons. 
        }
	\label{EFig_errthresh}
\end{figure}

\clearpage
\newpage
\FloatBarrier
\setcounter{figure}{0}
\captionsetup[figure]{labelfont={bf},name={Ext. Data Table},labelsep=bar,justification=raggedright,font=small}
\section*{Extended Data Tables}
\begin{figure*}[t!]
	\centering
	\includegraphics[width=\textwidth]{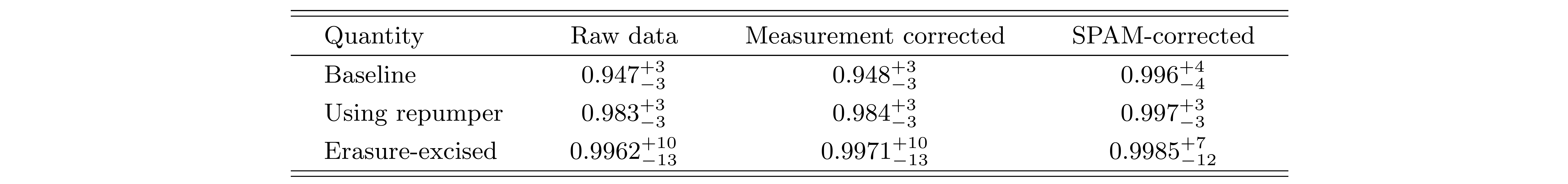}
        \caption{\textbf{Bell state fidelity lower bounds.}}
	\label{EFig:table_Bell}
\end{figure*}

\begin{figure}[t!]
	\centering
	\includegraphics[width=\columnwidth]{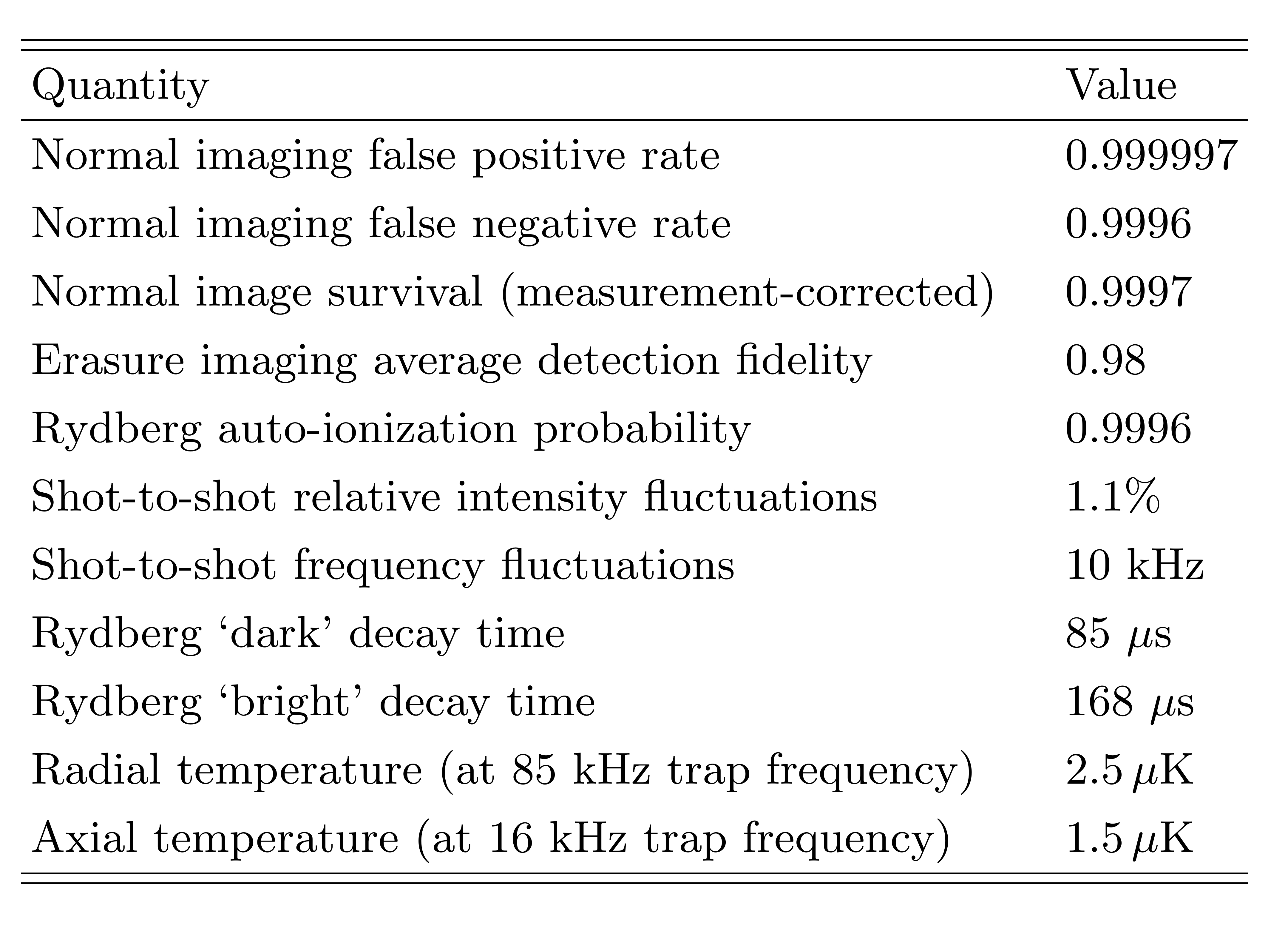}
        \caption{\textbf{Parameters of the error model.}}
	\label{EFig:table}
\end{figure}

\end{document}